# MAGNETO-OPTICS OF QUANTUM WIRES WITH $D^{(-)}$-CENTERS

**V.D. Krevchik[1], A.B. Grunin[1], A.K. Aringazin[2,3], M.B. Semenov[1,3], E.N. Kalinin[1], V.G. Mayorov[1], A.A. Marko[1], S.V. Yashin[1]**

[1]Department of Physics, Penza State University, Penza, 440017, Russia

physics@diamond.stup.ac.ru

[2] Institute for Basic Research, Department of Theoretical Physics,

Eurasian National University, Astana, 473021, Kazakhstan

aringazin@mail.kz

[3]Institute for Basic Research, P.O. Box 1577, Palm Harbor, Fl 34682, USA

ibr@gte.net

The impurity magneto-optical absorption for the cases of longitudinal and transversal light polarization with respect to the quantum wire axis has been theoretically studied. Analytical expressions for the corresponding $D^{(-)}$-centers photo-ionization cross sections under the action of longitudinal magnetic field have been obtained, and their spectral dependences for InSb quantum wires have been investigated.

## 1. Introduction

Magneto-optical properties of quantum wires (QW) attract increasing attention because of a number of interesting effects, which are related to the magnetic and dimensional quantizing hybridization [1], can be observed in these systems. Another interesting aspect is due to the fact that the magnetic field acting along the QW axis plays the role of a variable parameter, with the help of which an effective geometrical size of the system can be changed and, hence, optical transition energies can be controlled. This may be important in various applications, particularly, in designing photo-detectors with variable sensitivity.

The intraband magneto-optical absorption for quantum nanostructures with parabolic confinement potential has been theoretically studied [2]. In this paper, the appearance of the hybridization effect for optical absorption spectra in quantum ring with finite width, in QW and quantum cylinder, was demonstrated. Real quantum wires may contain impurity centers arising due to QW production technologies. Also, alloyage additions (impurities) can be made by using the $\delta$ -alloyage technology [3] in order to change physical properties of QW semiconductive structures. In this connection an interest to the problem of impurity centers binding energy governed modulation [4] in magnetic field, and, correspondingly, to the problem of low-dimensional alloyage-structures optical properties control, arises.

In this paper, we make a theoretical study of the magneto-optical impurity absorption in QW semiconductive structures, within the framework of the parabolic confinement potential model treated by zero-range potential method.

## 2. The D$^{(-)}$ -center energy spectrum in longitudinal magnetic field

Let us consider the positional disorder effect for semiconductive quantum wire (QW) with the parabolic confinement potential in longitudinal magnetic field. We will suppose, that the shape of QW is represented by the round cylinder with the radius $L$ considerably smaller than its length $L_z$ ($L << L_z$). To describe QW one-electron states, we will use the symmetric confinement potential,

$$V_1(\rho) = \frac{m^*}{2}\omega_0^2 \rho^2, \tag{1}$$

where $\rho \le L$, ($\rho, \varphi, z$) are cylindrical coordinates, $m^*$ is the electron effective mass, $\omega_0$ is the characteristic frequency of the QW confinement potential.

The external magnetic field is directed along the QW axis, $\vec{B} = (0, 0, B)$. The vector-potential $\vec{A}(\vec{r})$ can be chosen in the symmetric gauge,

$$\vec{A} = \frac{1}{2}\left[\vec{B}, \vec{r}\right], \tag{2}$$

that is, $\vec{A} = (-y\,B/2, x\,B/2, 0)$.

For unperturbed one-electron states in longitudinal magnetic field, the Hamiltonian of the chosen model in the cylindrical system of reference can be written as

$$H = -\frac{\hbar^2}{2m^*}\left(\frac{1}{\rho}\frac{\partial}{\partial\rho}\left(\rho\frac{\partial}{\partial\rho}\right) + \frac{1}{\rho^2}\frac{\partial^2}{\partial\varphi^2}\right) - \frac{i\hbar\omega_B}{2}\frac{\partial}{\partial\varphi} + \frac{m^*}{2}\left(\omega_0^2 + \frac{\omega_B^2}{4}\right)\rho^2 + H_z \tag{3}$$

where $\omega_B = |e|B/m^*$ is cyclotron frequency, $|\mathring{a}|$ is the electron charge, and $H_z = \left(-\hbar^2/(2m^*)\right)\partial^2/\partial z^2$.

Eigenvalues and eigenfunctions of the Hamiltonian (3) are found as [5]

$$E_{n,m,k}=\frac{\hbar\omega_B m}{2}+\hbar\omega_0\sqrt{1+\frac{\omega_B^2}{4\omega_0^2}}\left(2n+|m|+1\right)+\frac{\hbar^2k^2}{2m^*},\qquad (4)$$

$$\Psi_{n,m,k}(\rho,\varphi,z)=\frac{1}{2\pi a_1}\left[\frac{n!}{\left(n+|m|\right)!}\right]^{\frac{1}{2}}\left(\frac{\rho^2}{2a_1^2}\right)^{\frac{|m|}{2}}\exp\left(-\frac{\rho^2}{4a_1^2}\right)L_n^{|m|}\left(\frac{\rho^2}{2a_1^2}\right)\exp(im\varphi)\exp(ikz),\qquad (5)$$

where $n=0,1,2,...$ is the quantum number corresponding to Landau levels, $m=0,\pm1,\pm2,...$ is magnetic quantum number, $k$ is the electron quasi-momentum projection to the $z$-axis, $a_1^2=a^2/\left(2\sqrt{1+a^4/\left(4a_B^4\right)}\right)$, $a=\sqrt{\hbar/\left(m^*\omega_0\right)}$, $a_B=\sqrt{\hbar/\left(m^*\omega_B\right)}$ is magnetic length, and $L_n^c(x)$ is Laguerre polynomial [6].

It should be noted that in the used model the QW potential amplitude $U_0$ is an empirical parameter and, hence, the expressions (4) and (5) are fulfilled if

$$U_0\sqrt{1+\omega_B^2/\left(4\omega_0^2\right)}/\left(\hbar\omega_0\right)>>1,\qquad (6)$$

where $U_0=m^*\omega_0^2L^2/2$.

We suppose that the impurity center (IC, or $D^{(-)}$-center) is positioned at the point $\vec{R}_a=\left(\rho_a,\varphi_a,z_a\right)$, where $\rho_a,\varphi_a,z_a$ are the $D^{(-)}$-center cylindrical coordinates. The impurity potential has been described in terms of the zero-range potential with the intensity $\gamma$, $\gamma=2\pi\hbar^2/\left(\alpha m^*\right)$. This potential is of the following form:

$$V_\delta(\rho,\varphi,z;\rho_a,\varphi_a,z_a)=\gamma\frac{\delta(\rho-\rho_a)}{\rho}\delta(\varphi-\varphi_a)\delta(z-z_a)\left[1+(\rho-\rho_a)\frac{\partial}{\partial\rho}+(z-z_a)\frac{\partial}{\partial z}\right],\qquad (7)$$

where $\alpha$ can be determined by the binding energy $E_i$ at the same IC in massive semiconductor; $\delta(x)$ is Dirac delta-function.

It is known [7] that such a model can be used to describe the $D^{(-)}$-states corresponding to the additional electron joining to small donor. In the effective mass approximation, the wave-function, $\Psi_{\lambda_B}^{(QW)}(\rho,\varphi,z;\rho_a,\varphi_a,z_a)$, of the electron localized on short-range IC potential, satisfies the Schroedinger equation,

$$\left(E_{1\lambda_B}-H\right)\Psi_{\lambda_B}^{(QW)}(\rho,\varphi,z;\rho_a,\varphi_a,z_a)=V_\delta(\rho,\varphi,z;\rho_a,\varphi_a,z_a)\Psi_{\lambda_B}^{(QW)}(\rho,\varphi,z;\rho_a,\varphi_a,z_a), \quad (8)$$

where $E_{1\lambda_B}=-\hbar^2\lambda_B^2/\left(2m^*\right)$ are eigenvalues of the Hamiltonian $H_B^\delta=H+V_\delta(\rho,\varphi,z;\rho_a,\varphi_a,z_a)$. The subscript $B$ indicates dependence on the magnetic field induction $B$. The one-electron Green-function for Schroedinger equation (8), which corresponds to the source at the point $\vec{r}_1=(\rho_1,\varphi_1,z_1)$ and energy $E_{1\lambda_B}$, can be written as

$$G\left(\rho,\varphi,z,\rho_1,\varphi_1,z_1;E_{1\lambda_B}\right)=\int_{-\infty}^{+\infty}dk\sum_{n,m}\frac{\Psi_{n,m,k}^*(\rho_1,\varphi_1,z_1)\Psi_{n,m,k}(\rho,\varphi,z)}{\left(E_{1\lambda_B}-E_{n,m,k}\right)}. \quad (9)$$

The Lippman-Schwinger equation for the QW with parabolic potential profile $D^{(-)}$-state in longitudinal magnetic field can be represented as

$$\Psi_{\lambda_B}^{(QW)}(\rho,\varphi,z;\rho_a,\varphi_a,z_a)=\int_{-\infty}^{+\infty}\int_0^{2\pi}\int_0^{+\infty}\rho_1 d\rho_1 d\varphi_1 dz_1 G\left(\rho,\varphi,z,\rho_1,\varphi_1,z_1;E_{1\lambda_B}\right)\times$$
$$\times V_\delta(\rho_1,\varphi_1,z_1;\rho_a,\varphi_a,z_a)\Psi_{\lambda_B}^{(QW)}(\rho_1,\varphi_1,z_1;\rho_a,\varphi_a,z_a). \quad (10)$$

Substituting (7) into (10), we obtain

$$\Psi_{\lambda_B}^{(QW)}(\rho,\varphi,z;\rho_a,\varphi_a,z_a)=\gamma G\left(\rho,\varphi,z,\rho_a,\varphi_a,z_a;E_{1\lambda_B}\right)\times$$
$$\times\left(T_1\Psi_{\lambda_B}^{(QW)}\right)(\rho_a,\varphi_a,z_a;\rho_a,\varphi_a,z_a), \quad (11)$$

where

$$\left(T_1\Psi_{\lambda_B}^{(QW)}\right)(\rho_a,\varphi_a,z_a;\rho_a,\varphi_a,z_a)\equiv$$

$$\equiv \lim_{\substack{\rho\to\rho_a\\ \varphi\to\varphi_a\\ z\to z_a}}\left[1+(\rho-\rho_a)\frac{\partial}{\partial\rho}+(z-z_a)\frac{\partial}{\partial z}\right]\Psi_{\lambda_B}^{(QW)}(\rho,\varphi,z;\rho_a,\varphi_a,z_a). \qquad (12)$$

The action of the operator $T_1$ to both sides of (11) gives us the equation which determines the D$^{(-)}$-center binding state energy $E_{1\lambda_B}$ dependence on QW parameters, impurity positions $\vec{R}_a=(\rho_a,\varphi_a,z_a)$ and magnetic field $B$ value:

$$\alpha=\frac{2\pi\hbar^2}{m^*}(T_1G)\left(\rho_a,\varphi_a,z_a,\rho_a,\varphi_a,z_a;E_{1\lambda_B}\right). \qquad (13)$$

Using the one-particle wave-functions (5) and energies (4) in the Green function (11) we have

$$G(\rho,\varphi,z,\rho_a,\varphi_a,z_a;E_{1\lambda_B})=\frac{1}{4\pi^2a_1^2}\exp\left[-\frac{\rho_a^2+\rho^2}{4a_1^2}\right]\int_{-\infty}^{+\infty}dk\exp[ik(z-z_a)]\times$$
$$\times\sum_{n,m}\frac{n!}{(n+|m|)!}\left(\frac{\rho_a\rho}{2a_1^2}\right)^{|m|}L_n^{|m|}\left(\frac{\rho_a^2}{2a_1^2}\right)L_n^{|m|}\left(\frac{\rho^2}{2a_1^2}\right)\exp[im(\varphi-\varphi_a)]\times$$
$$\times\left(E_{1\lambda_B}-\frac{\hbar\omega_Bm}{2}-\hbar\omega_0\sqrt{1+\frac{\omega_B^2}{4\omega_0^2}}(2n+|m|+1)-\frac{\hbar^2k^2}{2m^*}\right)^{-1}. \qquad (14)$$

Let us consider the case when impurity level is positioned lower than the QW potential bottom $(E_{1\lambda_B}<0)$. We will rewrite the Green-function (14) in terms of the effective Bohr radius units ($a_d=4\pi\varepsilon_0\varepsilon\hbar^2/(m^*|e|^2)$ is the effective Bohr radius, where $\varepsilon_0$ is dielectric constant, $\varepsilon$ is the static relative dielectric permeability of QW semiconductive material), and the effective Bohr energy units, ($E_d=\hbar^2/(2m^*a_d^2)$ is the effective Bohr energy).

Calculation of the series sum over quantum number $n$ in Eq. (14) can be made with help of the Hille-Hardi formula for bilinear generating function [8],

$$\sum_{n=0}^{\infty}\frac{n!}{\Gamma(n+\alpha+1)}L_n^{\alpha}(x)L_n^{\alpha}(y)z^n=(1-z)^{-1}\exp\left(-z\frac{x+y}{1-z}\right)(xyz)^{-\frac{\alpha}{2}}I_{\alpha}\left(2\frac{\sqrt{xyz}}{1-z}\right),|z|<1. \tag{15}$$

Here, $I_{\alpha}(u)$ is the modified Bessel function of the first kind and $\Gamma(s)$ is Euler gamma-function.

Then, the Green-function (14) can be rewritten as

$$\begin{aligned}
G(\rho,\varphi,z,\rho_a,\varphi_a,z_a;E_{1\lambda_B}) = -\frac{1}{2^3\pi^{\frac{3}{2}}E_d a_d^3\sqrt{\beta}}\Bigg(\int_0^{+\infty}\frac{1}{\sqrt{t}}\exp\left[-\left((\beta\eta_{1B}^2+w)t+\frac{(z-z_a)^2}{4\beta a_d^2 t}\right)\right]\times \\
\times\left[2w\left(1-\exp[-2wt]\right)\right]^{-1}\times \\
\times\exp\left[-\frac{(\rho_a^2+\rho^2)w}{4\beta a_d^2}\frac{(1+\exp[-2wt])}{(1-\exp[-2wt])}\right]\times \\
\times\exp\left[\frac{1}{2}\left(\exp\left[i(\varphi-\varphi_a)-\beta a^{*-2}t\right]+\exp\left[-i(\varphi-\varphi_a)+\beta a^{*-2}t\right]\right)\times\right. \\
\left.\times\frac{\rho_a\rho\, w\exp[-wt]}{\beta a_d^2(1-\exp[-2wt])}\right]-\frac{1}{t}\exp\left[-\frac{(\rho-\rho_a)^2 w}{4\beta a_d^2 t}\right]\Bigg]dt+ \\
+2\sqrt{\pi\beta}\,a_d\frac{\exp\left[-\sqrt{\dfrac{(\beta\eta_{1B}^2+w)\left((\rho-\rho_a)^2 w+(z-z_a)^2\right)}{\beta a_d^2}}\right]}{\sqrt{(\rho-\rho_a)^2 w+(z-z_a)^2}}\Bigg).
\end{aligned} \tag{16}$$

Substituting (16) into (13), we obtain the following equation which determines the binding state energy $E_{1\lambda_B}$ dependence $(E_{1\lambda_B}<0)$ on D$^{(-)}$-center position $\vec{R}_a=(\rho_a,\varphi_a,z_a)$, QW parameters and magnetic field induction $B$ [9],

$$\sqrt{\eta_{1B}^2+\beta^{-1}w}=\eta_i-\frac{1}{\sqrt{\pi\beta}}\int_0^{+\infty}\frac{1}{\sqrt{t}}\exp\left[-\left(\beta\eta_{1B}^2+w\right)t\right]\times$$
$$\times\left(\frac{1}{2t}-w\left(1-\exp[-2wt]\right)^{-1}\exp\left[-\frac{\rho_a^{*\,2}w}{2\beta\left(1-\exp[-2wt]\right)}\times\right.\right.$$
$$\left.\left.\times\left(1+\exp[-2wt]-\left(\exp\left[-\beta a^{*^{-2}}t\right]+\exp\left[\beta a^{*^{-2}}t\right]\right)\exp[-wt]\right)\right]\right)dt, \qquad (17)$$

where $w=\sqrt{1+\beta^2 a^{*^{-4}}}$ , $\eta_i^2=\left|E_i\right|/E_d$ is the parameter which characterizes the binding state energy $E_i$ for the same impurity center in massive semiconductor, and $\rho_a^*=\rho_a/a_d$ .

Let us consider the case when impurity level $E_{1\lambda_B}$ is positioned between the two-dimensional oscillator potential well bottom (this potential well describes the QW potential), and the electron ground state energy level $E_{0,0,0}=\hbar\omega_0\sqrt{1+\omega_B^2/\left(4\omega_0^2\right)}$ in QW: $E_{1\lambda_B}=\hbar^2\lambda_B'^2/\left(2m^*\right)>0$; and the parameter $\eta_{1B}'^{\,2}=\lambda_B'^2 a_d^2$ can be introduced. Replacing $\lambda_B^2$ by $-\lambda_B'^2$ , or $\eta_{1B}^2$ by $-\eta_{1B}'^2$ leads to transition from the case $E_{1\lambda_B}<0$ to the case $E_{1\lambda_B}>0$. Hence, the transcendental equation, which determines the D$^{(-)}$-center binding energy $E_{1\lambda_B}$ dependence $\left(E_{1\lambda_B}>0\right)$ on QW parameters, impurity polar radius $\rho_a$ , and magnetic field induction $B$ value, becomes [9]

$$\sqrt{-\eta_{1B}'^2+\beta^{-1}w}=\eta_i-\frac{1}{\sqrt{\pi\beta}}\int_0^{+\infty}\frac{1}{\sqrt{t}}\exp\left[-\left(w-\beta\eta_{1B}'^2\right)t\right]\times$$
$$\times\left(\frac{1}{2t}-w\left(1-\exp[-2wt]\right)^{-1}\exp\left[-\frac{\rho_a^{*\,2}w}{2\beta\left(1-\exp[-2wt]\right)}\times\right.\right.$$
$$\left.\left.\times\left(1+\exp[-2wt]-\left(\exp\left[-\beta a^{*^{-2}}t\right]+\exp\left[\beta a^{*^{-2}}t\right]\right)\exp[-wt]\right)\right]\right)dt. \quad (18)$$

Because of the quantum dimensional effect the D$^{(-)}$-center binding energy $E_{\lambda_B}^{(QW)}$ for QW in longitudinal magnetic field should be determined as [10]

$$E_{\lambda_B}^{(QW)} = \begin{cases} E_{0,0,0} + \left|E_{1\lambda_B}\right|, & E_{1\lambda_B} < 0, \\ E_{0,0,0} - E_{1\lambda_B}, & E_{1\lambda_B} > 0 \end{cases} \quad (19)$$

or, in atomic units,

$$E_{\lambda_B}^{(QW)} / E_d = \begin{cases} \beta^{-1} w + \eta_{1B}^2, & E_{1\lambda_B} < 0, \\ \beta^{-1} w - \eta_{1B}'^2, & E_{1\lambda_B} > 0, \end{cases} \quad (20)$$

where $E_{0,0,0}$ is determined due to Eq. (4).

The two possible cases in Eq. (19) correspond to impurity level position lower ($E_{1\lambda_B} < 0$) and higher ($E_{1\lambda_B} > 0$) than the QW bottom. Fig. 1 represents the results of numerical analysis of Eq. (18) for semiconductive QW D$^{(-)}$-states (based on InSb); the effective mass of electron in InSb and the static relative dielectric permeability are: $m^* = 0.0133 m_0$ (where $m_0$ is the electron rest mass), and $\varepsilon = 18$, correspondingly; and the effective Bohr energy is $E_d \approx 5.5 \times 10^{-4}\ eV$. As one can see from Fig. 1, in both the cases $E_{1\lambda_B} > 0$ and $E_{1\lambda_B} < 0$ (curves 1 and 2, respectively) the binding energy of D$^{(-)}$-center is a decreasing function of the radial coordinate $\rho_a$, that is related with the dimensional quantizing. The D$^{(-)}$-center binding energy $E_{\lambda_B}^{(QW)}$ considerably increases in the presence of magnetic field (see curves 3 and 4 of Fig. 1). In the case $E_{1\lambda_B} < 0$, the binding energy increases (as one can see by comparing curves 2 and 4) by more than $0.02\, eV$ for D$^{(-)}$-centers, which are placed on the QW-axis. Thus, existence conditions for a bound state in longitudinal magnetic field becomes less restrictive (as one can see by comparing curves 1 and 3, 2 and 4 of Fig. 1). We conclude that the magnetic field stabilizes the QW D$^{(-)}$-states.

The possibility to control D$^{(-)}$-centers ionization energy by the magnetic field allows one to change the charge carriers concentration in rather wide range because of exponential dependence of the distribution function on the energy near the QW Fermi level. The wave-function of electron which is localized on the $D^{(-)}$-center short-range potential in QW with parabolic potential profile in longitudinal magnetic field, as one can see from Eq. (11), differs only by a factor from the one-electron Green-function. The Green-function (16) is then written as

$$G\left(\rho, \varphi, z, \rho_a, \varphi_a, z_a; E_{1\lambda_B}\right) = \frac{1}{a^3 \hbar \omega_0} G^{(1)}\left(\rho, \varphi, z, \rho_a, \varphi_a, z_a; E_{1\lambda_B}\right), \tag{21}$$

where $G^{(1)}\left(\rho, \varphi, z, \rho_a, \varphi_a, z_a; E_{1\lambda_B}\right)$ is the dimensionless Green-function. Therefore, in accord to Eq. (11), for the wave-function $\Psi_{\lambda_B}^{(QW)}\left(\rho, \varphi, z; \rho_a, \varphi_a, z_a\right)$ we obtain

$$\Psi_{\lambda_B}^{(QW)}\left(\rho, \varphi, z; \rho_a, \varphi_a, z_a\right) = -C_B G^{(1)}\left(\rho, \varphi, z, \rho_a, \varphi_a, z_a; E_{1\lambda_B}\right), \tag{22}$$

where $C_B$ is the normalization constant,

$$C_B = 2^{\frac{3}{4}} \pi^{\frac{1}{2}} \beta^{-\frac{3}{4}} a_d^{-\frac{3}{2}} w^{\frac{1}{4}} \left[\zeta\left(\frac{3}{2}, \frac{\beta \eta_{1B}^2}{2w} + \frac{1}{2}\right)\right]^{-\frac{1}{2}}, \tag{23}$$

so that the binding state wave-function is obtained as follows:

$$\Psi_{\lambda_B}^{(QW)}\left(\rho, \varphi, z; z_a\right) = 2^{\frac{1}{4}} \pi^{-1} \beta^{-\frac{3}{4}} a_d^{-\frac{3}{2}} w^{\frac{5}{4}} \left[\zeta\left(\frac{3}{2}, \frac{\beta \eta_{1B}^2}{2w} + \frac{1}{2}\right)\right]^{-\frac{1}{2}} \times$$
$$\times \int_0^{+\infty} \frac{1}{\sqrt{t}} \exp\left(-\left[\left(\beta \eta_{1B}^2 + w\right)t + \frac{\left(z - z_a\right)^2}{4\beta a_d^2 t}\right]\right) \left(1 - \exp[-2wt]\right)^{-1} \times$$
$$\times \exp\left[-\frac{\rho^2 w}{4\beta a_d^2} \frac{\left(1 + \exp[-2wt]\right)}{\left(1 - \exp[-2wt]\right)}\right] dt, \tag{24}$$

where $\Psi_{\lambda_B}^{(QW)}\left(\rho, \varphi, z; z_a\right) \equiv \Psi_{\lambda_B}^{(QW)}\left(\rho, \varphi, z; 0, 0, z_a\right)$.

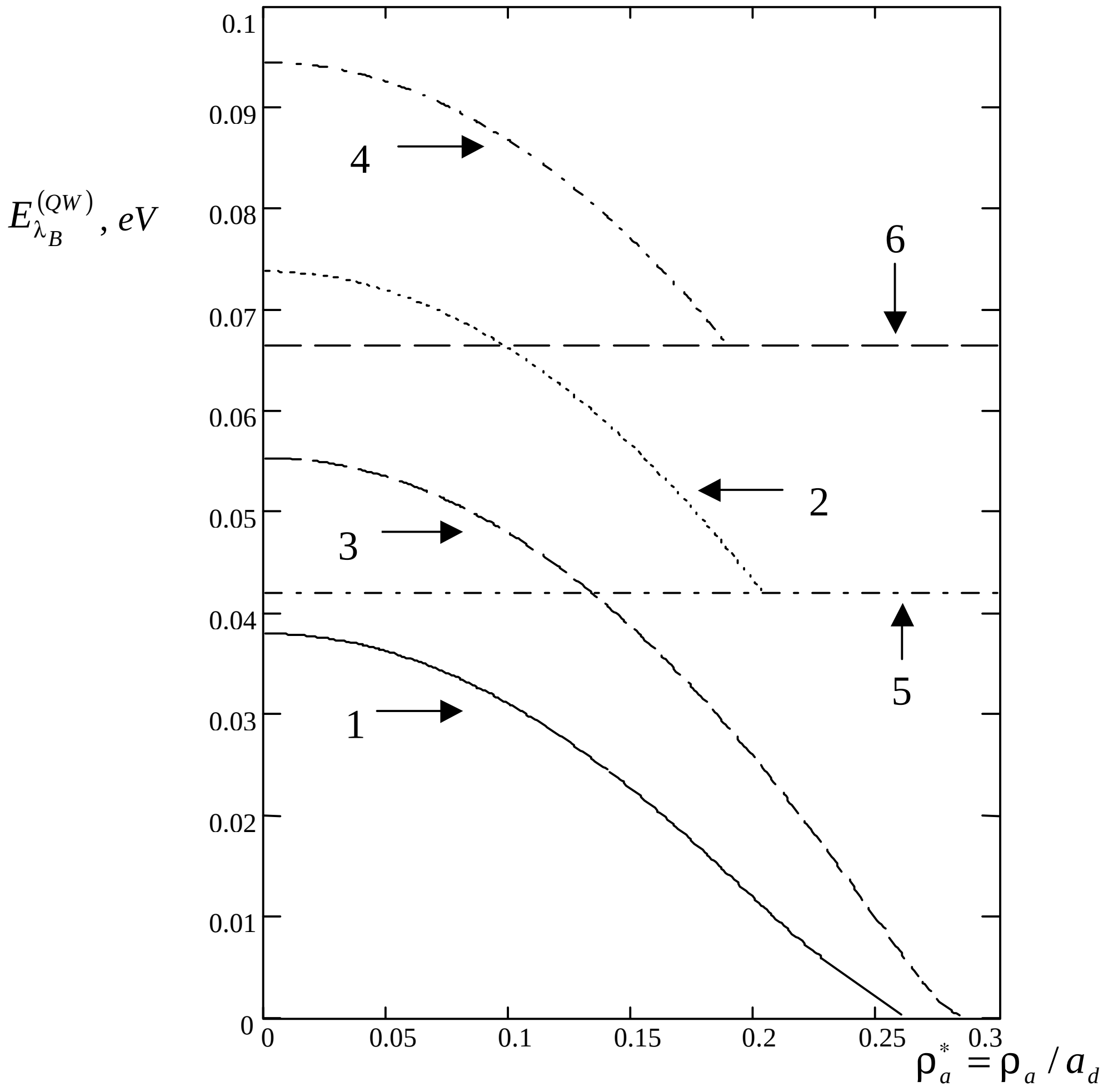

Fig. 1. The D$^{(-)}$-center binding energy $E_{\lambda_B}^{(QW)}$ dependence, (for QW based on InSb), from polar IC-radius $\rho_a^* = \rho_a / a_d$ for different values of magnetic induction $B$; (curves 1 and 3 correspond to the case $E_{1\lambda_B} > 0$, curves 2 and 4 correspond to the case $E_{1\lambda_B} < 0$; the energy levels positions in the QW ground state for $B = 0\,T$ and $B = 12\,T$ are depicted by curves 5 and 6, respectively; $L$ = 35.8 *nm*, $U_0$ = 0.2 *eV*): 1 – $|E_i| = 5\times10^{-3}\ eV$, $B = 0\,T$; 2 – $|E_i| = 3.5\times10^{-2}\ eV$, $B = 0\,T$; 3 – $|E_i| = 5\times10^{-3}\ eV$, $B = 12\,T$; 4 – $|E_i| = 3.5\times10^{-2}\ eV$, $B = 12\,T$.

## 3. D$^{(-)}$ -centers photo-ionization section in longitudinal magnetic field

Let us consider the light impurity absorption for QW with parabolic potential profile in longitudinal magnetic field (along the QW-axis): $\vec{B}\uparrow\uparrow\vec{k}$, $\vec{B}$ is the magnetic induction vector, and $\vec{k}$ is the *z*-axis unit vector. It is supposed that impurity center is localized at the point $\vec{R}_a=(0,0,0)$ and the D$^{(-)}$ –center binding state energy level $E_{1\lambda_B}$ is lower than the QW potential bottom, $\left(E_{1\lambda_B}<0\right)$. In accordance with Eq. (24), the wave-function for the QW D$^{(-)}$ –center binding state in longitudinal magnetic field is chosen as follows:

$$\Psi_{\lambda_B}^{(QW)}(\rho,\varphi,z;0)=2^{\frac{1}{4}}\pi^{-1}\beta^{-\frac{3}{4}}a_d^{-\frac{3}{2}}w^{\frac{5}{4}}\left[\zeta\left(\frac{3}{2},\frac{\beta\eta_{1B}^2}{2w}+\frac{1}{2}\right)\right]^{-\frac{1}{2}}\times$$
$$\times\int_0^{+\infty}\frac{1}{\sqrt{t}}\exp\left(-\left[\left(\beta\eta_{1B}^2+w\right)t+\frac{z^2}{4\beta a_d^2 t}\right]\right)\left(1-\exp[-2wt]\right)^{-1}\times$$
$$\times\exp\left[-\frac{\rho^2 w}{4\beta a_d^2}\frac{\left(1+\exp[-2wt]\right)}{\left(1-\exp[-2wt]\right)}\right]dt\,. \qquad (25)$$

Since we consider the impurity electron strong localization case,

$$\sqrt{2}\lambda_B a_1>>1, \qquad (26)$$

where $\lambda_B{}^2\equiv 2m^*\left|E_{1\lambda_B}\right|/\hbar^2$, the finite state wave-function can be chosen in the form (5).

The effective Hamiltonian $H_{\mathrm{int}\,B}$ for interaction with the light wave field in the longitudinal magnetic field (along the QW-axis) can be written as

$$H_{\mathrm{int}\,B}=-i\hbar\lambda_0\sqrt{\frac{2\pi\hbar^2\alpha^*}{m^{*2}\omega}}I_0\exp\left(i\vec{q}\,\vec{r}\right)\left(\left(\vec{e}_\lambda\,\nabla_{\vec{r}}\right)-\frac{i|e|B}{2\hbar}\left[\vec{e}_\lambda,\vec{r}\right]_z\right), \qquad (27)$$

where $\lambda_0=E_{eff}/E_0$ is the local field coefficient, which accounts for the optical transition amplitude increase because of the fact that the D$^{(-)}$ –center effective local field $E_{eff}$ exceeds the mean macroscopic field $E_0$ in crystal;

$\alpha^{*}=|e|^{2}/\left(4\pi\varepsilon_{0}\sqrt{\varepsilon}\,\hbar c\right)$ is the fine structure constant with account for the static relative dielectric permeability $\varepsilon$; $c$ is the light speed in vacuum; $I_0$ is the light intensity; $\omega$ is the frequency for absorbed radiation with the wave-vector $\vec{q}$ and the polarization unit vector $\vec{e}_{\lambda}$; $\nabla_{\vec{r}}$ is the Hamiltonian operator; $|e|$ is the charge of electron; $B$ is the magnetic field induction absolute value.

For case of absorption for the longitudinal polarization light (in relation to the QW-axis), $\vec{e}_{\lambda s}=(0,0,1)$, the effective Hamiltonian (27) is of the form

$$H^{(s)}_{\mathrm{int}\,B}=-i\hbar\lambda_0\sqrt{\frac{2\pi\hbar^{2}\alpha^{*}}{m^{*2}\omega}I_0}\,\exp\left(i\vec{q}_s\vec{r}\right)\left(\vec{e}_{\lambda s}\nabla_{\vec{r}}\right), \qquad (28)$$

where the superscript $s$ denotes the electromagnetic wave longitudinal polarization.

Matrix elements $M^{(s)}_{f,\lambda_B}$, which determine the oscillator force value for the electron dipole optical transitions from the IC ground state $\Psi^{(QW)}_{\lambda_B}(\rho,\varphi,z;0)$ to the QW quasi-discrete spectrum states $\Psi_{n,m,k}(\rho,\varphi,z)$ in magnetic field for the longitudinal light polarization $\vec{e}_{\lambda s}$ case, are written as

$$M^{(s)}_{f,\lambda_B}=i\lambda_0\sqrt{\frac{2\pi\alpha^{*}I_0}{\omega}}\left(E_{n,m,k}-E_{1\lambda_B}\right)\left\langle\Psi^{*}_{n,m,k}(\rho,\varphi,z)\right|\left(\vec{e}_{\lambda s},\vec{r}\right)\left|\Psi^{(QW)}_{\lambda_B}(\rho,\varphi,z;0)\right\rangle. \qquad (29)$$

With the account for the charge carriers energetic spectrum expression (4) and the finite state wave-function $\Psi_{n,m,k}(\rho,\varphi,z)$, Eq. (5), as well as the IC binding state wave-function $\Psi^{(QW)}_{\lambda_B}(\rho,\varphi,z;0)$, Eq. (25), for matrix elements $M^{(s)}_{f,\lambda_B}$ we get

$$M^{(s)}_{f,\lambda_B}=2^{-\frac{1}{4}-\frac{|m|}{2}}\,i\pi^{-\frac{3}{2}}\lambda_0\sqrt{\frac{\alpha^{*}I_0}{\omega}}\beta^{-\frac{9}{4}-\frac{|m|}{2}}w^{\frac{7}{4}+\frac{|m|}{2}}E_d\,a_d^{-|m|-\frac{5}{2}}\left[\zeta\left(\frac{3}{2},\frac{\beta\eta^{2}_{1B}}{2w}+\frac{1}{2}\right)\right]^{-\frac{1}{2}}\times$$

$$\times\left[\frac{n!}{(n+|m|)!}\right]^{\frac{1}{2}}\left(m\beta a^{*-2}+w\left(2n+|m|+1\right)+\beta k^{2}a_d^{2}+\beta\eta^{2}_{1B}\right)\int_0^{+\infty}\int_{-\infty}^{+\infty}\int_0^{2\pi}d\varphi\,dz\,d\rho\,\rho^{|m|+1}\times$$

$$\times\exp\left[-\frac{\rho^2 w}{4\beta a_d^2}\right]L_n^{|m|}\left(\frac{\rho^2 w}{2\beta a_d^2}\right)\exp(-im\varphi)\exp(-ikz)\times z\times$$

$$\times\int_0^{+\infty}dt\frac{1}{\sqrt{t}}\exp\left[-\left(\beta\eta_{1B}^2+w\right)t\right]\exp\left[-\frac{z^2}{4\beta a_d^2 t}\right]\left(1-\exp[-2wt]\right)^{-1}\times$$

$$\times\exp\left[-\frac{\rho^2 w}{4\beta a_d^2}\frac{\left(1+\exp[-2wt]\right)}{\left(1-\exp[-2wt]\right)}\right]. \quad (30)$$

When calculating the matrix elements $M_{f,\lambda_B}^{(s)}$ for the considered optical transitions, the following integrals were used [11]:

$$\int_0^{2\pi}\exp(-im\varphi)d\varphi=\begin{cases}0, & m\neq 0,\\ 2\pi, & m=0,\end{cases}$$

(31)

$$\int_0^{+\infty}\rho\exp\left[-\frac{\rho^2 w}{2\beta a_d^2\left(1-\exp[-2wt]\right)}\right]L_n\left(\frac{\rho^2 w}{2\beta a_d^2}\right)d\rho=$$

$$=\frac{\beta a_d^2}{w}\left(1-\exp[-2wt]\right)\exp[-2nwt]. \quad (32)$$

Integration over $z$ coordinate leads to the following expression [11]:

$$\int_{-\infty}^{+\infty}z\exp\left[-\frac{z^2}{4\beta a_d^2 t}-ikz\right]dz=-4i\sqrt{\pi}\beta^{\frac{3}{2}}a_d^3 kt^{\frac{3}{2}}\exp\left[-\beta a_d^2 k^2 t\right]. \quad (33)$$

As one can see from (31), the selection rule for the magnetic quantum number $m$ allow optical transitions from impurity level only to the QW states with $m=0$. Hence, taking into account of Eqs. (33), (31) and (32), for the matrix elements (30) we obtain

$$M_{f,\lambda_B}^{(s)}=2^{\frac{11}{4}}\lambda_0\sqrt{\frac{\alpha^* I_0}{\omega}}\beta^{\frac{1}{4}}w^{\frac{3}{4}}E_d a_d^{\frac{5}{2}}\left[\zeta\left(\frac{3}{2},\frac{\beta\eta_{1B}^2}{2w}+\frac{1}{2}\right)\right]^{-\frac{1}{2}}\frac{k}{\left(\beta\eta_{1B}^2+\beta a_d^2 k^2+(2n+1)w\right)}.$$

(34)

The photo-ionization section $\sigma_B^{(s)}(\omega)$ for the impurity center, which is localized at the point $\vec{R}_a = (0,0,0)$, in the longitudinal polarization $\vec{e}_{\lambda s}$ light absorption case, is determined as

$$\sigma_B^{(s)}(\omega) = \frac{2\pi}{\hbar I_0} \int_{-\infty}^{+\infty} dk \sum_n \left| M_{f,\lambda_B}^{(s)} \right|^2 \delta\left( E_{n,0,k} + \left| E_{1\lambda_B} \right| - \hbar\omega \right). \qquad (35)$$

The integral in Eq. (35) requires to find roots $k^{(1),(2)}$ of the argument of the Dirac delta-function which satisfies the energy conservation law for electron optical transitions from IC binding state to the QW-states as result of photon absorption with $\hbar\omega$ energy:

$$\beta^{-1} w(2n+1) + k^2 a_d^2 + \eta_{1B}^2 - X = 0, \qquad (36)$$

where $X = \hbar\omega / E_d$ is photon energy in the effective Bohr energy $E_d$ units.

The roots $k^{(1),(2)}$ for Eq. (36) are:

$$k^{(1),(2)} = \pm a_d^{-1} \sqrt{X - \eta_{1B}^2 - \beta^{-1} w(2n+1)}. \qquad (37)$$

Using Eqs. (34) and (37), the photo-ionization section $\sigma_B^{(s)}(\omega)$ expression (35) can be written as

$$\sigma_B^{(s)}(\omega) = \sigma_0 \beta^{-\frac{3}{2}} w^{\frac{3}{2}} \left[ \zeta\left( \frac{3}{2}, \frac{\beta\eta_{1B}^2}{2w} + \frac{1}{2} \right) \right]^{-1} X^{-3} \times$$

$$\times \sum_{n=0}^{N} \theta\left( X - \eta_{1B}^2 - \beta^{-1} w(2n+1) \right) \sqrt{X - \eta_{1B}^2 - \beta^{-1} w(2n+1)}, \qquad (38)$$

where $\sigma_0 = 2^{13/2} \pi \alpha^* \lambda_0^2 a_d^2$; $N = [C_2]$ is the integer part of the number $C_2 = \beta(X - \eta_{1B}^2)/(2w) - 1/2$; $\theta(x)$ is the Heaviside unit-step function [12],

$$\theta(x) = \begin{cases} 1, & x \geq 0, \\ 0, & x < 0. \end{cases} \qquad (39)$$

The photon cutoff energy $X_{thB}^{(s)}$ (in the Bohr units) for the light impurity absorption case with longitudinal polarization (in relation to QW-axis) $\vec{e}_{\lambda s}$, can be found due to the equation

$$X_{th_B}^{(s)} = \eta_{1B}^2 + \beta^{-1} w , \qquad (40)$$

where $X_{thB}^{(s)} = \hbar\omega_{thB}^{(s)} / E_d$, $\hbar\omega_{thB}^{(s)}$ is the photon cutoff energy in usual units. Fig. 2 shows the photo-ionization section spectral dependence $\sigma_B^{(s)}(\omega)$ for impurity center with $\vec{R}_a = (0, 0, 0)$ in the light longitudinal polarization case for QW, (which based on InSb). As one can see from Fig. 2, the photo-ionization section is characterized by a non-monotonic spectral dependence, which is associated to the double quantization. Besides, the oscillations period equals $\hbar\Omega$, ($\Omega = \sqrt{4\omega_0^2 + \omega_B^2}$), i.e. it is determined by the hybrid frequency $\Omega$. A comparison of the curves 1 and 2 shows that the impurity absorption band edge in magnetic field is shifted to the short-wave spectrum region.

Fig. 3 represents the cutoff energy $\hbar\omega_{thB}^{(s)}$ dependence for the longitudinal polarization photon in the light impurity absorption case (for QW based on InSb) upon the magnetic induction value *B*. As it follows from Fig. 3, the impurity absorption band edge monotonously increases with the increase of the magnetic field, and its shift at magnetic field induction value *B* = 12 *T* is bigger than 0.02 *eV*.

Let us consider light absorption by IC $\left(\vec{R}_a = (0,0,0)\right)$ in QW with parabolic potential profile in longitudinal magnetic field in the case when the photon wave-vector $\vec{q}$ is directed along the QW-axis (the polarization vector $\vec{e}_\lambda$ is perpendicular to *Oz*-axis). In accordance with Eq. (27), the effective Hamiltonian $H_{\text{int}\,B}^{(t)}$ of interaction with the light wave field, for the transversal polarization $\vec{e}_{\lambda t}$ case (in relation to QW-axis) in longitudinal magnetic field, can be written as

$$H_{\text{int}\,B}^{(t)} = -i\hbar\lambda_0 \sqrt{\frac{2\pi\hbar^2\alpha^*}{m^{*2}\omega} I_0} \exp\left(i\vec{q}_t\,\vec{r}\right)\left(\left(\vec{e}_{\lambda_t}\,\nabla_{\vec{r}}\right) - \frac{i|e|B}{2\hbar}\left[\vec{e}_{\lambda_t}, \vec{r}\right]_z\right), \qquad (41)$$

where $\vec{q}_t = (0, 0, q_z)$ is the photon wave-vector and $q_z$ is the wave-vector projection $\vec{q}_t$ to the *z*-axis.

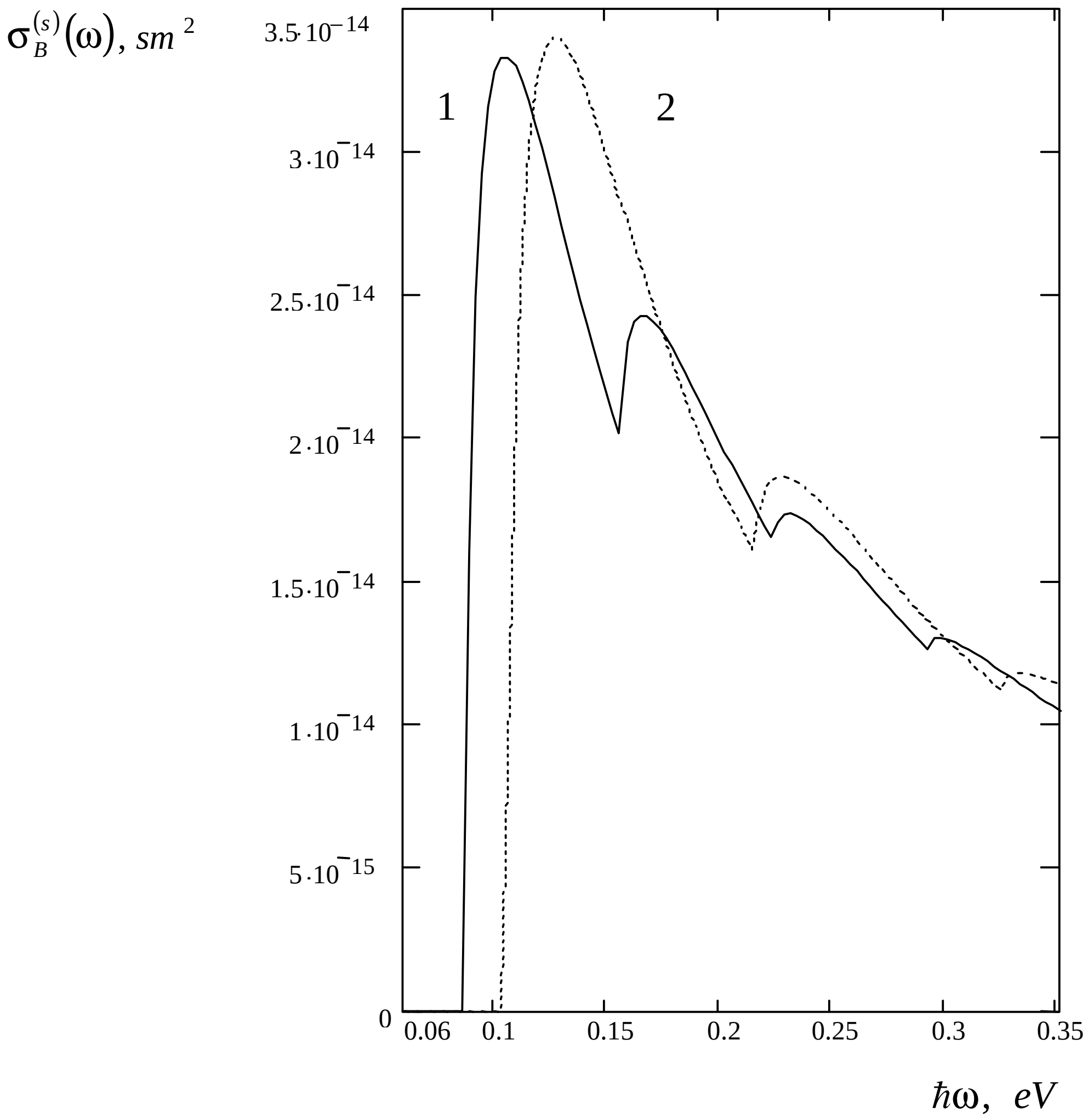

Fig. 2. The $D^{(-)}$- center photo-ionization section $\sigma_B^{(s)}(\omega)$ spectral dependence for QW based on InSb (in the light longitudinal polarization case, $\vec{R}_a = (0,0,0)$, $|E_i| = 5.5\times10^{-2}$ *eV*, *L*=53.7 *nm*, $U_0$=0.3 *eV*): curve 1 : $B = 0T$ ; curve 2 : $B = 10T$ .

Let us represent Eq. (41) in the cylindrical frame of coordinates,

$$H_{\text{int}\,B}^{(t)} = -i\hbar\lambda_0\sqrt{\frac{2\pi\hbar^2\alpha^*}{m^{*2}\omega}I_0}\exp\left(iq_z z\right)\times$$

$$\times\left(\cos\left(\theta-\varphi\right)\frac{\partial}{\partial\rho}+\frac{1}{\rho}\sin\left(\theta-\varphi\right)\frac{\partial}{\partial\varphi}-\frac{i|e|B}{2\hbar}\rho\sin\left(\varphi-\theta\right)\right), \qquad (42)$$

where $\theta$ is polar angle of the transversal polarization unit vector $\vec{e}_{\lambda t}$ in the cylindrical frame. In the dipole approximation the electron-photon interaction matrix elements $M_{f,\lambda_B}^{(t)}$, which determine electron transitions from $D^{(-)}$ – center ground state to QW – states as the result of the absorption of photon with polarization $\vec{e}_{\lambda t}$, can be written as

$$M_{f,\lambda_B}^{(t)} = \left\langle\Psi_{n,m,k}^*\left(\rho,\varphi,z\right)\middle|H_{\text{int}\,B}^{(t)}\middle|\Psi_{\lambda_B}^{(QW)}\left(\rho,\varphi,z;0\right)\right\rangle =$$

$$= -i\hbar\lambda_0\sqrt{\frac{2\pi\hbar^2\alpha^* I_0}{m^{*2}\omega}}\int_0^{2\pi}\int_{-\infty}^{+\infty}\int_0^{+\infty}\rho\,d\rho\,d\varphi\,dz\,\Psi_{n,m,k}^*\left(\rho,\varphi,z\right)\times$$

$$\times\left(\cos\left(\theta-\varphi\right)\frac{\partial}{\partial\rho}+\frac{1}{\rho}\sin\left(\theta-\varphi\right)\frac{\partial}{\partial\varphi}-\frac{i|e|B}{2\hbar}\rho\sin\left(\varphi-\theta\right)\right)\Psi_{\lambda_B}^{(QW)}\left(\rho,\varphi,z;0\right). \qquad (43)$$

Using the one-electron states in longitudinal magnetic field (5) and wave-function (25) for the IC binding state in QW, the matrix elements $M_{f,\lambda_B}^{(t)}$ can be evaluated as follows:

$$M_{f,\lambda_B}^{(t)} = 2^{-\frac{1}{4}-\frac{|m|}{2}}i\pi^{-\frac{3}{2}}\lambda_0\sqrt{\frac{\alpha^* I_0}{\omega}}E_d a_d^{-|m|-\frac{1}{2}}\beta^{-\frac{|m|}{2}-\frac{5}{4}}w^{\frac{|m|}{2}+\frac{7}{4}}\left[\zeta\left(\frac{3}{2},\frac{\beta\eta_{1B}^2}{2w}+\frac{1}{2}\right)\right]^{-\frac{1}{2}}\left[\frac{n!}{\left(n+|m|\right)!}\right]^{\frac{1}{2}}\times$$

$$\times\int_0^{+\infty}dt\,\frac{1}{\sqrt{t}}\exp\left[-\left(\beta\eta_{1B}^2+w\right)t\right]\left(1-\exp\left[-2wt\right]\right)^{-1}\times$$

$$\times\int_0^{+\infty}d\rho\,\rho^{|m|+2}\exp\left[-\frac{\rho^2 w}{2\beta a_d^2\left(1-\exp\left[-2wt\right]\right)}\right]L_n^{|m|}\left(\frac{\rho^2 w}{2\beta a_d^2}\right)\times$$

$$\times\int_{-\infty}^{+\infty}dz\,\exp\left[-ikz-\frac{z^2}{4\beta a_d^2 t}\right]\int_0^{2\pi}\exp\left(-im\varphi\right)\left(\cos\left(\theta-\varphi\right)\frac{w}{\beta a_d^2}\times\right.$$

$$\left.\times\frac{\left(1+\exp\left[-2wt\right]\right)}{\left(1-\exp\left[-2wt\right]\right)}+\frac{i}{a_B^2}\sin\left(\varphi-\theta\right)\right)d\varphi\,, \qquad (44)$$

where $a_B$ – is the magnetic length.

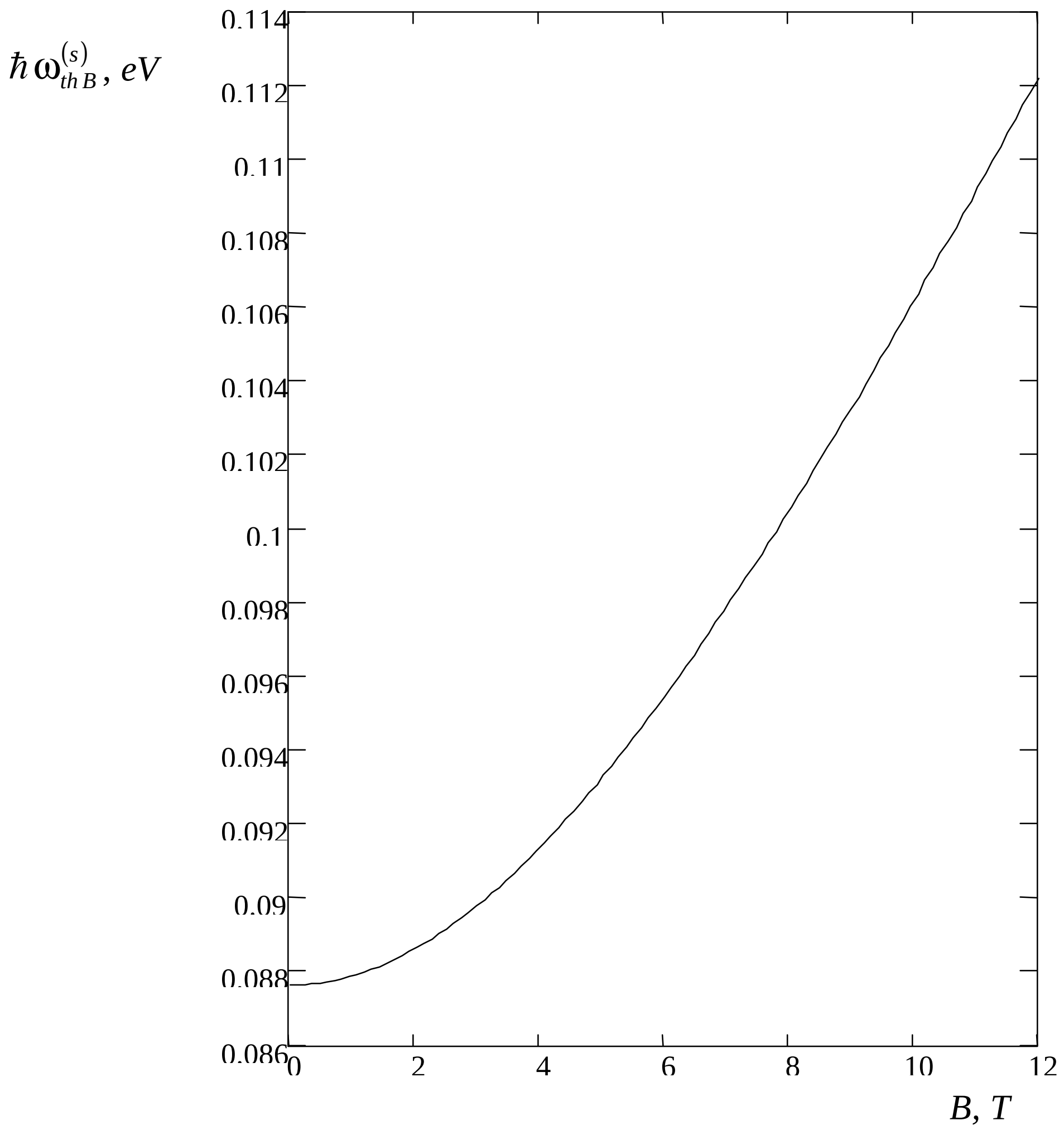

Fig. 3. The photon cutoff energy $\hbar\omega_{thB}^{(s)}$ dependence, for the longitudinal polarization light impurity absorption case in QW based on InSb ($\vec{R}_a = (0,0,0)$, $|E_i| = 5.5\times10^{-2}$ *eV*, *L*=53.7 *nm*, $U_0$=0.3 *eV*), on the magnetic induction *B*.

Estimation of the matrix elements (44) leads to calculation of the integral [11], which determines the magnetic quantum number $m$ selection rule,

$$\int_0^{2\pi}\exp(-im\varphi)\left(\cos(\theta-\varphi)\frac{w}{\beta a_d^2}\frac{(1+\exp[-2wt])}{(1-\exp[-2wt])}+\frac{i}{a_B^2}\sin(\varphi-\theta)\right)d\varphi=$$

$$=\pi\exp(\mp i\,\theta)\delta_{m,\,\pm1}\left(\frac{w}{\beta a_d^2}\frac{(1+\exp[-2wt])}{(1-\exp[-2wt])}+\frac{m}{a_B^2}\right). \qquad (45)$$

Here $\delta_{m,\,\pm1}$ is the Kronecker symbol,

$$\delta_{m,\,\pm1}=\begin{cases}1, & m=\pm1,\\ 0, & m\neq\pm1,\end{cases} \qquad (46)$$

with the sign «–» in the exponent index $\exp(\mp i\,\theta)$ corresponds to $m=+1$, and the sign «+» to $m=-1$, correspondingly. One can see from Eqs. (45) and (46) that the optical transitions from impurity level can occur only to QW-states with the quantum number $m=\pm1$. Evaluation of Eq. (44) requires the integral [11]

$$\int_{-\infty}^{+\infty}\exp\left[-\frac{z^2}{4\beta a_d^2 t}-i\,k\,z\right]dz=2\sqrt{\pi\beta}\,a_d\sqrt{t}\,\exp\left[-\beta a_d^2\,k^2 t\right] \qquad (47)$$

and the estimation, with account for the magnetic quantum number selection rule [11],

$$\int_0^{+\infty}\rho^3\exp\left[-\frac{\rho^2 w}{2\beta a_d^2(1-\exp[-2wt])}\right]L_n^1\left(\frac{\rho^2 w}{2\beta a_d^2}\right)d\rho=$$

$$=\frac{2\beta^2 a_d^4}{w^2}(n+1)(1-\exp[-2wt])^2\exp[-2nwt]. \qquad (48)$$

Using Eqs. (45), (46), (47) and (48) for matrix elements $M_{f,\lambda_B}^{(t)}$ we finally obtain

$$M_{f,\lambda_B}^{(t)} = 2^{\frac{9}{4}} i \lambda_0 \exp(\mp i\theta)\delta_{m,\pm 1}\sqrt{\frac{\alpha^* I_0}{\omega}} E_d a_d^{\frac{3}{2}} \beta^{-\frac{1}{4}} w^{\frac{5}{4}} \left[\zeta\left(\frac{3}{2}, \frac{\beta\eta_{1B}^2}{2w} + \frac{1}{2}\right)\right]^{-\frac{1}{2}} (n+1)^{\frac{1}{2}} \times$$

$$\times \frac{\left(\beta\eta_{1B}^2 + (2n+2)w + m\beta a^{*-2} + \beta a_d^2 k^2\right)}{\left(\beta\eta_{1B}^2 + (2n+1)w + \beta a_d^2 k^2\right)\left(\beta\eta_{1B}^2 + (2n+3)w + \beta a_d^2 k^2\right)}. \quad (49)$$

The photo-ionization section $\sigma_B^{(t)}(\omega)$ for IC, which is placed at the point $\vec{R}_a = (0,0,0)$, for the photons (with wave-vector $\vec{q}_t$ and polarization vector $\vec{e}_{\lambda t}$) absorption case, can be calculated due to the following formula:

$$\sigma_B^{(t)}(\omega) = \frac{2\pi}{\hbar I_0} \int_{-\infty}^{+\infty} dk \sum_n \sum_{m=-1}^{1} \left| M_{f,\lambda_B}^{(t)} \right|^2 \delta\left(E_{n,m,k} + \left|E_{1\lambda_B}\right| - \hbar\omega\right). \quad (50)$$

Roots of the argument of $\delta$- function are found from

$$m a^{*-2} + \beta^{-1} w \left(2n + |m| + 1\right) + k^2 a_d^2 + \eta_{1B}^2 - X = 0, \quad (51)$$

where $m = \pm 1$. The roots are

$$k_*^{(1),(2)} = \pm a_d^{-1} \sqrt{X - \eta_{1B}^2 - m a^{*-2} - \beta^{-1} w \left(2n + |m| + 1\right)}. \quad (52)$$

Using Eqs. (49) and (52), we write the photo-ionization section,

$$\sigma_B^{(t)}(\omega) = \sigma_0 \times 2^{-1} \beta^{\frac{3}{2}} w^{\frac{5}{2}} \left[\zeta\left(\frac{3}{2}, \frac{\beta\eta_{1B}^2}{2w} + \frac{1}{2}\right)\right]^{-1} X \times$$

$$\times \sum_{n=0}^{N'} (n+1) \sum_{m=-1}^{1} \delta_{m,\pm 1} \theta\left(X - \eta_{1B}^2 - m a^{*-2} - \beta^{-1} w \left(2n + |m| + 1\right)\right) \times$$

$$\times\left(X-\eta_{1B}^2-m\,a^{*-2}-\beta^{-1}w\left(2n+|m|+1\right)\right)^{-\frac{1}{2}}\times$$
$$\times\left[\beta^2\left(X-(m+1)a^{*-2}\right)\left(X-(m-1)a^{*-2}\right)-1\right]^{-2}, \qquad (53)$$

where $\sigma_0=2^{13/2}\pi\alpha^*\lambda_0^2\,a_d^2$; $N^/=[C_3]$ is an integer part of the number

$$C_3=\beta\left(X-\eta_{1B}^2+a^{*\,-2}\right)/(2\,w)-1,$$

$\theta(x)$ is the Heaviside unit-step function (39).

The photon cutoff energy $X_{th\,B}^{(t)}=\hbar\,\omega_{th\,B}^{(t)}/E_d$ (in Bohr units) for case of the light impurity absorption with transversal (in relation to QW-axis) polarization $\vec{e}_{\lambda t}$, is determined by the relation,

$$X_{th\,B}^{(t)}=\eta_{1B}^2+2\beta^{-1}w-a^{*\,-2}, \qquad (54)$$

where $\hbar\,\omega_{th\,B}^{(t)}$ is the photon cutoff energy in usual units.

Fig. 4 shows the spectral dependence for IC photo-ionization section $\sigma_B^{(t)}(\omega)$ with $\vec{R}_a=(0,0,0)$ in the light transversal polarization case for QW based on InSb. As one can see from Fig. 4, the spectrum for the transversal polarization light magneto-optical impurity absorption is a series of resonance peaks with a doublet structure. The distance between the resonance doublet peaks equals to $\hbar\,\omega_B$, i.e., it is determined by the cyclotron frequency $\omega_B$. Doublets are positioned periodically at absorption curve with the period $\hbar\Omega$. Resonant frequencies are determined by the general formula,

$$\omega_{res}=\left|E_{1\lambda_B}\right|/\hbar+m\omega_B/2+\Omega\,(2n+|m|+1)/2\ ,$$

and essentially depend on the impurity level position depth and upon the magnetic field intensity.

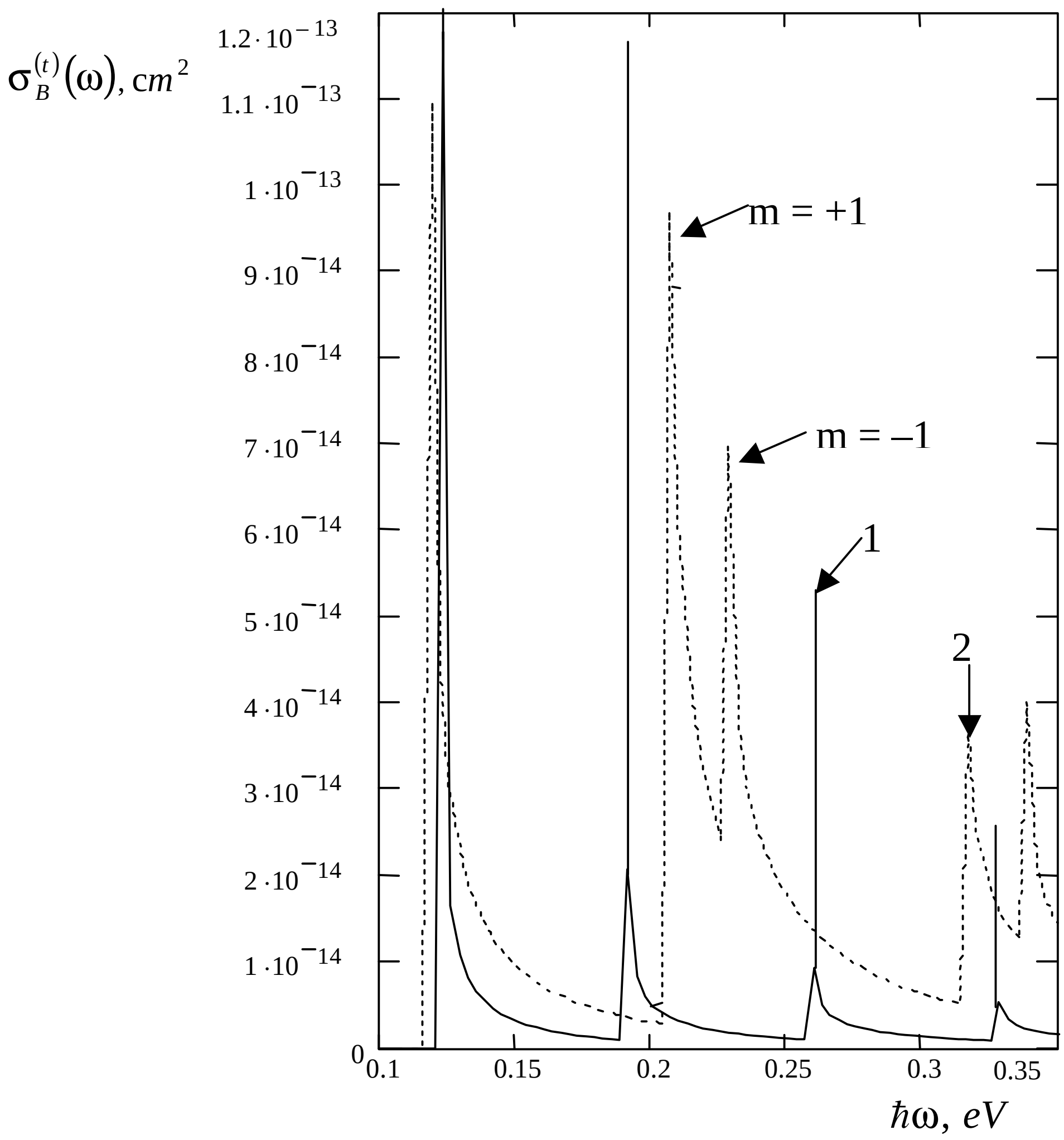

Fig. 4. The $D^{(-)}$–center photo-ionization section $\sigma_B^{(t)}(\omega)$ spectral dependence in QW based on InSb, (for the light transversal polarization case, $\vec{R}_a = (0,0,0)$, $|E_i| = 5.5\times10^{-2}$ eV, *L*=53.7 *nm*, $U_0$=0.3 *eV*); curve 1: $B = 0\,T$ ; curve 2: $B = 10\,T$ .

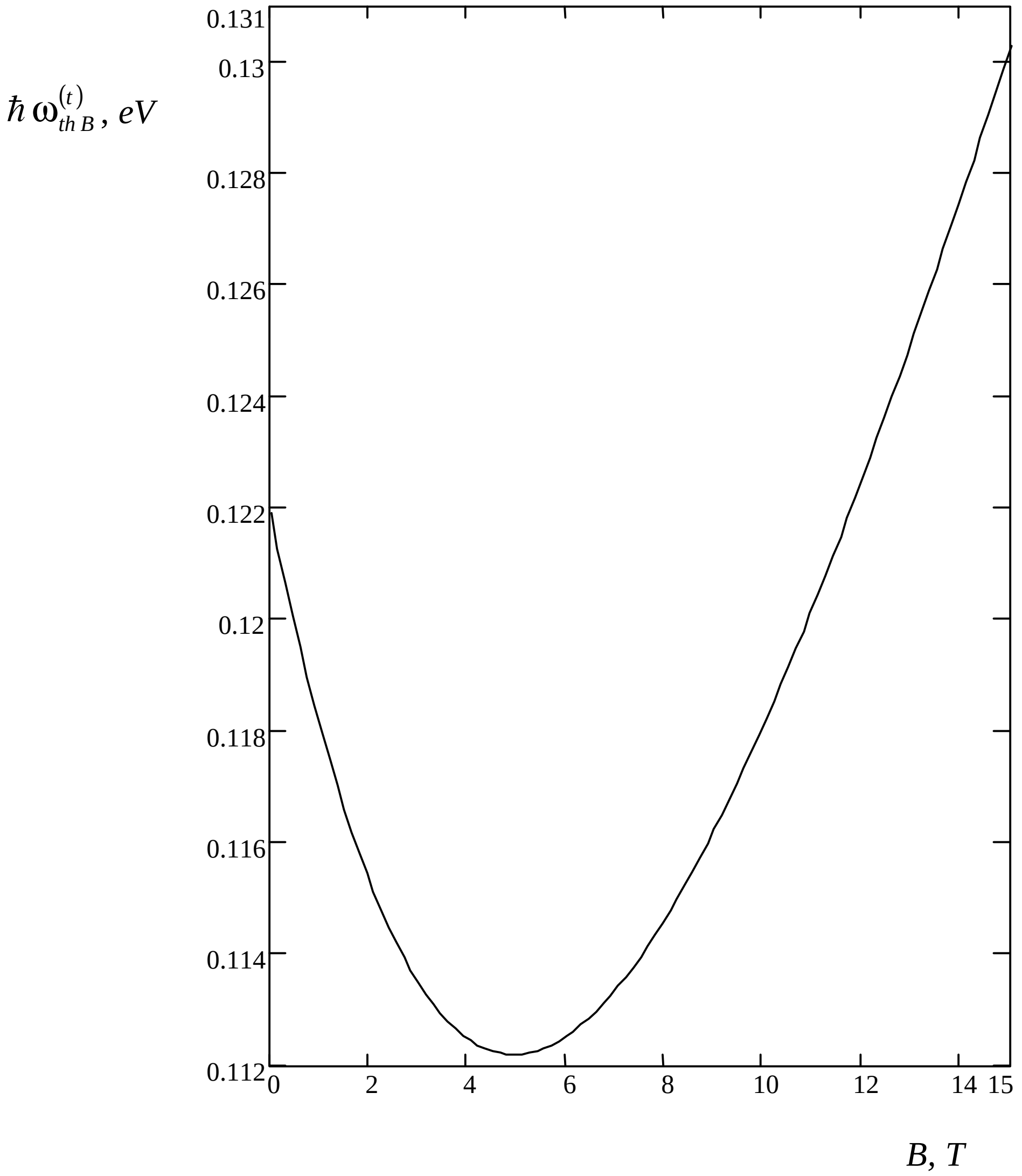

Fig. 5. The photon cutoff energy $\hbar\,\omega_{th\,B}^{(t)}$ dependence, for the light (with transversal polarization) impurity absorption case in QW based on InSb ($\vec{R}_a = (0,0,0)$, $|E_i| = 5.5\times 10^{-2}$ $eV$, $L$=53.7 *nm*, $U_0$ = 0.3 *eV*), on the magnetic induction value $B$ .

In Fig. 5 we represent the photon cutoff energy $\hbar\omega_{th\,B}^{(t)}$ dependence, for the case of the light (transversal polarization) impurity absorption in InSb QW, on the magnetic induction intensity *B*. One can see that the displayed dependence is of a non-monotonic character with pronounced minimum.

## 4. Conclusions

In the present paper, the local impurity states for semiconductive QW with parabolic confinement potential in longitudinal magnetic field have been theoretically studied. Within the framework of the zero-range potential model and the effective mass approximation, the bound states problem for QW with $D^{(-)}$–center" in magnetic field has been treated analytically, with exact results being obtained. It has been shown that the influence of magnetic field leads to appreciable variation of positions of the impurity levels and to a stabilization of the bound states in QW. It has been found that the spectrum for the transversal polarization light magneto-optical impurity absorption is a series of the resonance peaks with a doublet structure. It was also found that the doublet peaks are placed one from the other at the distance which is determined by cyclotron frequency; and doublets are positioned periodically with period equal to the hybrid frequency.